\documentclass[pra,superscriptaddress,twocolumn,showpacs,footnotes]{revtex4}

\begin{document}

\title{Unexpected reemergence of von Neumann theorem}

\author{Marek \.Zukowski}
\affiliation{Institute of Theoretical Physics and Astrophysics, University
of Gda\'nsk, ul. Wita Stwosza 57, PL-80-952 Gda\'nsk, Poland}
%\affiliation{National Centre for Quantum Information of Gda\'nsk, ul. W. Andersa 27, PL-81-824 Sopot, Poland}

\pacs{03.65.Ud}

\begin{abstract}
 Is is shown here that the  ``simple test of quantumness for a single system'' of arXiv:0704.1962  (for a recent experimental realization see arXiv:0804.1646) has exactly the same relation to the discussion of to the problem of describing the quantum system via a classical probabilistic scheme (that is in terms of hidden variables, or within a realistic theory) as the von Neumann theorem (1932). The latter one was shown by Bell (1966) to stem from an assumption that the hidden variable values for a sum of two non-commuting observables (which is an observable too) have to be, for each individual system, equal to sums of eigenvalues of the two operators. One cannot find a physical justification for such an assumption to hold for non-commeasurable variables. On the positive side. the criterion may be useful in rejecting models which are based on stochastic classical fields. Nevertheless the example used by the Authors has a classical optical realization.     
\end{abstract}
\date{\today}

\maketitle

\section{Introduction}
The no-go theorem for classical probabilistic (i.e., hidden variable) description of single quantum systems of ref.\cite{ALICKI1}, 
called by the Authors {\em a simple test of quantumness for a single system},  reads
\begin{itemize}
\item
If one can find two {\em non-negative} observables $A$ and $B$, such that $B-A$ is a {\em non-negative} operator too, and a state for which the averages of these observables and their squares
have the following property $0\leq\langle A\rangle_{avg} \leq \langle B\rangle_{avg}$ and $\langle A^2\rangle_{avg} > \langle B^2\rangle_{avg}$, then these measurements on the system are not describable in  terms of ``minimal classical models''.
\end{itemize}
This the formulation of the test is of ref. \cite{ALICKI2}, in which the claim is scaled down a bit. Also, the Authors admit in \cite{ALICKI2} that the criterion does not test {\em general hidden variable models}, but rather ``eliminated a well-defined classical theory for some specific systems''.

While one cannot the challenge the logical value of this thesis, the following question can be put: which specific hidden variable models are excluded by the criterion?
 It will be shown here that the criterion  of \cite{ALICKI1,ALICKI2} excludes the same class of hidden variable models as the von Neumann theorem \cite{VON} of 1932. The latter one was shown  in ref.\cite{BELL1966} to be too restrictive in its assumptions to be useful in the discussion of whether one can find classical probabilistic models of quantum mechanical processes. That is, it will be shown that minimal classical models are equivalent to von Neumann's assumptions on ``admissible'' hidden variables. This is very troubling, as it was shown that these assumptions are doubtful \cite{BELL1966}, \cite{MERMIN}.
 
Thus the usefulness of the criterion in the discussion of the foundations of quantum mechanics is highly limited. Nevertheless, it may be useful in pinpointing phenomena which have no description in terms of stochastic classical fields. 
However, the example given in \cite{ALICKI1, ALICKI2} and realized in \cite{EXP} does have a classical model, like every second order (in terms of the fields) photonic interference effect. The observed phenomena can be interpreted as non-classical only due to the statistical 
properties of the parametric down conversion process.  

\section{Relation with the von Neumann theorem }
{\em Any real linear combination of any two Hermitian operators} [say $A$ and $B$] {\em represents an observable, and the same linear combination of its expectation values is the expectation value of the combination} (quotation after Bell \cite{BELL1966}). This is innocent and true in quantum mechanics, but if one, following von Neumann, assumes that the same rule must hold also for all ``dispersion free states'' (i.e. deterministic classical models, the average over which should give the quantum averages), this immediately transfers this rule to the possible experimental results. That is for the hidden  dispersion free states von Neumann
tacitly assumed that $$v(A+B)=v(A)+v(B),$$ even if the quantum observables do not commute (are noncommeasurable). In more physical terms, if one has a system governed by hidden variables, then the allowed pure (classical) states are such that the above rule holds in each individual run of an experiment. But as the very conditions to measure $A+B$, $A$ and $B$ are different, see footnote \footnote{Because none of the pairs of out of these three observables commutes. In the laboratory this implies  different operational situations.},  thus there is no reason whatsoever to assume this. Therefore the von Neumann no-go theorem is inconsequential, \cite{MERMIN}.

All that was written above can be found in \cite{BELL1966} or \cite{MERMIN}. 
The Authors of \cite{ALICKI1} and \cite{ALICKI2} base their reasoning on the following theorem:
$0\leq A\leq B \Rightarrow A^2\leq B^2$ holds for all A and B belonging to an algebra $\textsl{A}$ if and only if the algebra $\textsl{A}$ is commutative (i.e. isomorphic to the algebra of continuous functions on a certain compact space).
Therefore the task now is to show that the assumptions of von Neumann when applied to two observables $A$ and $B$ are sufficient and necessary for the following:
if for all states $\rho$ one has $0\leq Tr(\rho A)\leq Tr(\rho  B)$ then $Tr(\rho A^2)\leq Tr(\rho B^2)$. 

The sufficiency derivation can be done by starting with a variant of the assumption of von Neumann $$v(B-A)=v(B)-v(A).$$ Since the eigenvalues of $B-A$ are non-negative, so must be $v(B-A)$ as they are always equal to one of the eigenvalues. 
Thus $v(B)\geq v(A)$ for each individual hidden variable steered system (or a dispersion free state). The non-negativity of $A$ and $B$ implies that $v(A)\geq0$ and $v(B)\geq0$. This obviously,  implies that for an individual system $v(B^2)=[v(B)]^2> v(A^2)=[v(A)]^2$. The equations stem again from the quantum rules concerning eigenvalues of {\em commuting} observables, whereas the inequality has only an algebraic origin. Obviously this implies that after averaging over dispersion free states, or series of individual realizations, one must always have that for triples $A\geq0$, $B\geq0$ and $B-A\geq0$
one has
\begin{eqnarray}
&\langle A\rangle_{avg} \leq \langle B\rangle_{avg}&\nonumber \\ &implies& \nonumber\\ &\langle A^2\rangle_{avg} \leq \langle B^2\rangle_{avg}.&\label{RULE}
\end{eqnarray}

The next task is to prove that the above rule, (\ref{RULE}), implies the von Neumann assumptions. This can be done using the very theorem on the $C^*$ algebras that the Authors of \cite{ALICKI1, ALICKI2} use to get their result. If $\langle A\rangle_{avg} < \langle B\rangle_{avg}$ implies $\langle A^2\rangle_{avg} < \langle B^2\rangle_{avg}$ for all possible states, then the operators $A$ and $B$ are from a commutative algebra. Once we have commutativity, obviously the von Neumann assumptions are true for pairs of observables $A$ and $B$. There is no problem with commeasurability, and even pure {\em quantum} states which are dispersion free for these observables (i.e. their eigenstates) satisfy the von Neumann assumption $v(B-A)=v(B)-v(A).$

\section{Non-minimal models}
Of course one should end this comment  giving an explicit counterexample - that is a hidden variable model that reproduces the predictions of the example given in ref. \cite{ALICKI1, ALICKI2}. Of course such a model will not be ``minimal''. However, this will not be put here as such a model can be found in \cite{EXP}, see footnote \footnote{However, the Authors of \cite{EXP} do not discuss all implications of the existence of such a model.}. 
But, as a matter of fact already in the famous work of Bell \cite{BELL1964} one can find an explicit hidden variable model for all (von Neumann) measurements on spin $1/2$ (qubit). 

Therefore one can ask the following question concerning the interpretation of the mathematical result of \cite{ALICKI1, ALICKI2}. If one can find two observables $A\geq0$ and $B\geq0$, such that $B-A\geq0$, and a state for which: \begin{eqnarray} 
&\langle A\rangle_{avg} \leq \langle B\rangle_{avg}&\nonumber \\ & and & \label{RULE2} \\ &\langle A^2\rangle_{avg} > \langle B^2\rangle_{avg},&\nonumber \\  \nonumber 
\end{eqnarray}
what is the implication of this property for the question of existence of classical probability
models of such averages? Definitely this not a general impossibility of a classical probabilistic model for these two observables, at least for one qubit (because of the existence of the a aforementioned model of ref. \cite{EXP}). However, it is very easy to show that one can have a plethora of hidden variable models with such a property. Simply, as it will be shown below,  the conjunction of these inequalities can be achieved if the observable A is governed by a hidden variable $\lambda_1$ whereas the observable B is governed by an independent hidden variable $\lambda_2$. Then the inequality (1) of \cite{ALICKI1, ALICKI2} does not apply, see footnote \footnote{The inequality is $0\leq A(x) \leq B(x)$, where, in the language of this comment, $x $ is a hidden variable. This is an assumption of a specific distinguishing property of the minimal models, which is not derivable from the properties of the observed the average $\langle A-B\rangle_{avg}.$}, and it is definitely  not impossible that
$$\int A(\lambda_1)\varrho_1(\lambda_1)d\lambda_1 \leq \int B(\lambda_2)\varrho_2(\lambda_2)d\lambda_2$$
and
$$\int [A(\lambda_1)]^2\varrho_1(\lambda_1)d\lambda_1 >\int [B(\lambda_2)]^2\varrho_2(\lambda_2)d\lambda_2,$$
while the joint hidden variable distribution is given by $$\varrho(\lambda_1,\lambda_2)=\varrho_1(\lambda_1)\varrho_2(\lambda_2).$$ 
Of course such a model is not minimal anymore.

One can always built such a model for a given pair of quantum observables, for any quantum state $\rho$, by the following construction (which is given here for $d$-state systems):
\begin{itemize}
\item
denote the eigenvalues of $A$ and $B$ by $A_i$ and $B_j$, respectively, with $i,j=1,2,...,d$, and
calculate the quantum probabilities for the given state for getting these results, $P(X_k|\rho)$ with $X=A, B$ and $k=1,2,...,d$,
\item
put for the hidden (variables) probability: $P_{HV}(A_i, B_j)= P(A_i|\rho)P(B_j|\rho)$, see footnote \footnote{Here the measurement outcomes, $X_k$, play the role of hidden variables.}. 
\end{itemize}

Obviously such a construction is universal, and thus it applies also to  pairs of non-negative quantum observables which have the property $Tr[\rho(B-A)]\geq 0 $ for all $\rho$. The hidden probabilities $P_{HV}$ reproduce correctly the quantum predictions for $A^2$ and $B^2$
for all states even if $Tr\rho(A^2-B^2)\geq 0$. For different  state preparations $\rho$ we have a different $P_{HV}$, but this is allowed even in classical physics. The model involves many hidden variables, but this is no problem. {One can always have as many hidden variables as one wishes, because they are hidden anyway. } 

Thus, the condition of \cite{ALICKI1, ALICKI2} is not a condition of a genuine quantumness, but rather one concludes that: if the condition (\ref{RULE2}) holds, then  there is no chance to have a  classical model in which for every individual system (or  dispersion free state) one can put $v(A-B)=v(A)-v(B)$, that is for which the von Neumann assumption is valid.

\subsection{Minimality loophole}
Finally, a remark is due. In the published version of their work \cite{ALICKI2} the Authors scale down their claim
to the following: the condition (\ref{RULE2}), if satisfied, prohibits {\em a minimal classical model} for the observables. Thus, the thesis of this work can be reduced to the following: minimal classical models of the published version of \cite{ALICKI1, ALICKI2} are equivalent to the von Neumann assumptions, and thus face the criticism which can be found in \cite{BELL1966} and \cite{MERMIN}.  The aim of this comment is to warn the readers about this, especially when attempting to interpret experimental data.

Let us now address the discussion of the ``minimality loophole'', ref. \cite{ALICKI2} page 3. One can read:
``One could still argue that there may exist a non-minimal classical AM [algebraic model] describing the
data ('minimality loophole'). In this case, the classical observable $B - A$ possesses negative
outcomes (values of the function) which are not detectable by the differences of averages
$\langle B\rangle - \langle A\rangle$.'' However, for non-commuting $A$ and $B$, since $B-A$ is not commeasurable with neither $A$, nor $B$, and even the latter ones are noncommeasurable too, there is no reason for the eigenvalues  to follow the von Neumann rule (as the Authors suggest). Thus  $v(B)-v(A)$ may be negative for individual system, while $v(B-A)\geq0$. One can 
explicitly construct a non-minimal model without negative $v(B-A)$:    
\begin{itemize}
\item
denote the eigenvalues of $A$, $B$ and $B-A=C$ are $A_i$, $B_j$, $C_k$, respectively, with $i,j,k=1,2,...,d$,
and
calculate the quantum probabilities for the given state for getting these results, $P(X_k|\rho)$ with $X=A, B, C$ and $k=1,2,...,d$,
\item
put for the hidden (variables) probability: $P_{HV}(A_i, B_j, C_k)= P(A_i|\rho)P(B_j|\rho)P(C_k|\rho)$. 
\end{itemize}

\section{Well defined classical models}
As far as a {\em direct} detection of non existence of {\it any} classical probabilistic models is concerned we are left with the two theorems of Bell, see \cite{MERMIN}, which do not use the von Neumann assumptions to limit the hidden variable theories. The two theorems are based, except for realism, on non-contextuality assumption \cite{BELL1966}, or the locality assumption \cite{BELL1964}. The latter one is based on a very strong relativistically motivated criterion of direct causal independence of events which are spatially separated (they can have a common cause, but cannot influence each other directly), and the additional `natural' assumption that one may have stochastic processes which are statistically independent (for details see, \cite{GILL}).    

However, it would be interesting to find a useful realm of applicability of the criterion of \cite{ALICKI1, ALICKI2}. The authors write that the criterion rule out {\em minimal} classical description. Such a description was shown above to be equivalent to the von Neumann theorem, but does this make it useless like the theorem? Certainly not. It ill be shown here that the condition makes impossible a classical probabilistic description which is using the tools of classical field theory, or aims at describing the system via phase space methods (e.g, Wigner quasi-distributions), as suggested by the Authors.
Simply, under the condition \cite{ALICKI1, ALICKI2}, an attempt to use such methods would fail - because the quasi-classical description requires the possibility of having either singular and/or non-positive distributions.  
Thus the form of non-classicality detected by the criterion in e.g. quantum optics is limited to the phenomena which do not have a model in terms of statistical distributions of random classical electromagnetic fields. 

To illustrate this, let us use the example given by the authors, and embed it into quantum optics of a single mode field.
This can easily be done e.g. by assuming that the sole two eigenstates of the operator $A$ are $|0\rangle$ and $|1\rangle$, i.e the vacuum state and the on photon state. Thus the eigenstates of the operator $B$ are some linear combinations of the two. So is the pure state $|\varphi> = 0.391|0\rangle + 0.920|1\rangle$,  which gives the 
optimal realization of the criterion: $\langle B^2\rangle>\langle A^2\rangle$ despite $A\geq B\geq 0$. 

A direct comparison with classical theory can be made if one uses the $P$ representation. It based on the overcomplete, continuous basis of coherent states, denoted here as $|\alpha\rangle$). In this formalism a general density operator reads
\begin{equation}
 \rho=\int P(\alpha)|\alpha\rangle\langle\alpha|d^2\alpha,
\end{equation}
with $d^2\alpha=d(Re\alpha)d(Im\alpha).$
  The operators are determined by their diagonal matrix elements, e.g. $\langle \alpha A| \alpha\rangle $, which are usually denoted as
  $A_Q(\alpha, \alpha^*)$.
The averages are given by
\begin{equation}
 Tr A\rho=\int A_Q(\alpha, \alpha^*)P(\alpha)|\alpha\rangle\langle\alpha|d^2\alpha. \label{TRACE}
\end{equation}
That is the operators are represented by functions, and therefore the picture apparently looks classical.
However, non classical phenomena occur when $P(\alpha)$
are not  non-negative, or are more singular than a delta function.
It is well known that superpositions of two Fock states is highly singular, thus suitably selected observables must reveal the impossibility of treating the relation (\ref{TRACE})
as a model employing probabilistic distribution of classical amplitudes, and a functional representation of the observables.

Thus the criterion of \cite{ALICKI1, ALICKI2} belongs to he same class of non-classical effects as antibunching or
 $100\%$ interferometric contrast of products of intensity fluctuations observed by multiple detectors, etc. More general claims based on the criterion are unfounded.   

These remarks on the example of the Authors given above were tailored such that the best features of the criterion were stressed.
However, one can also present a `complementary' discussion of the specific example given in \cite{ALICKI1, ALICKI2}. It will be shown that  the example has not only a classical model, but even a classical realization. Let s take a balanced Mach-Zehnder interferometer. It is well known that such a device is capable of performing any unitary ${\cal U}(2)$ transformation on a photonic qubit (with the two distinguishable states of being in the upper beam and in the lower beam). But it also  well known, that there is no distinction between second order interference (in terms of fields) in the quantum and classical realm, see footnote \footnote{In simpler words there is no distinction in the Young-type interference between classical fields and those revealing quantum nature.}.
Thus if one takes as the inputs to the two  entry ports of the interferometer, $1$ and $2$,  two classical analytic signals $I(t)_i=a_iI(t)$ with amplitudes, $a_i$, as in $|\varphi>$ (and both following the same temporal behavior, $I(t)$), then in the output 
ports one receives $a'_iI(t)$, with $a'_i=\sum_{i,j=1,2} U_{ij}a_j$ (we skip the retardation effects). The response of a  detector is proportional to the intensity of the field impinging on it. Thus, the probability to register a count in output $i$ is given by $<<|a'_iI(t)|^2>>$, where $<<>>$ denotes some time integration over the detector's time resolution.    

Therefore one can tune the interferometer in such a way that it unitarily transforms qubits of amplitude $(1,0)$ and $(0,1)$
into two basis states of the observable $A$, i.e.,  it performs a transformation $U(A)_{ij}$ (the other tunings will be for  $B$, and  $B-A$) of the example), and take as measurement results weighted averages of the counts a
the two detectors,  that is $$\langle A\rangle_{avg}=\frac{\sum_{i=1,2}\lambda(A)_i<<|a'_iI(t)|^2>>}{<<|I(t)|^2>>},$$ 
where $\lambda(A)_i$ are the weights which are equal to the eigenvalues of the observable $A$ of the example and $a'_i$ now stans for $\sum_{i,j=1,2} U_{ij}(A)a_j$ (and a similar formula for $B$ and $B-A$).  Just a glance reveals, that the predictions of the quantum example and this classical models are identical.
Even $\langle A-B\rangle_{avg}=\frac{\sum_{i=1,2}\lambda(A-B)_i<<|a'_iI(t)|^2>>}{<<|I(t)|^2>>}$ is always positive\footnote{One could argue here that one might split the incoming classical beams into two pairs and send them into two differently tuned Mach-Zehnder interferometers, and then observe fluctuations giving negative values, but such an experiment does not model the two output situation assumed in the example.}.

The same algebra holds in for a polarization version of the experiment (which is the case of \cite{EXP}). Non-classicality may be shown   only if one considers that photonic nature of light may introduce antibunching effects at the detection stations, or 
non-classical correlations between the trigger (idler) and the detectors measuring the polarization of the signal photon (as it was one in \cite{EXP}. Thus in in the experiment \cite{EXP} the sole non-classicality is due to the statistical  properties of the parametric down conversion process\footnote{Which, by the way, enables the heralding employed in the experiment.}, and cannot be analyzed directly using  the example of \cite{ALICKI1,ALICKI2}. In short, the test inapplicable in the ``single particle'' case (compare \cite{EXP}).

This work is a part of the EU 6-th Framework programme SCALA, and has been done at {\em National Centre for Quntum Information of Gdansk}. It has been finished before the appearance of \cite{ALICKI-07}

\end{document}